\documentclass[preprint,12pt,authoryear]{elsarticle}

\makeatletter
\def\ps@pprintTitle{%
 \let\@oddhead\@empty
 \let\@evenhead\@empty
 \def\@oddfoot{\centerline{\thepage}}%
 \let\@evenfoot\@oddfoot}
\makeatother

\usepackage{amssymb}
\usepackage{makecell}
\usepackage{epigraph} 
\usepackage{hyperref}
\usepackage{placeins} 
\usepackage{eurosym}

\journal{Journal of Behavioral and Experimental Finance}

\begin{document}

\begin{frontmatter}



\title{Take the aTrain. Introducing an Interface for the \textbf{A}ccessible \textbf{Tra}nscription of \textbf{In}terviews.}


\author[inst1]{Armin Haberl}

\affiliation[inst1]{organization={Business Analytics and Data Science-Center, University of Graz},
            addressline={Universitätsstraße 15}, 
            city={Graz},
            postcode={8010}, 
            country={Austria}}
\affiliation[inst2]{organization={Know Center \& Graz University of Technology},
            addressline={Sandgasse 36}, 
            city={Graz},
            postcode={8010}, 
            country={Austria}}
\author[inst1]{Jürgen Fleiß}
\author[inst2]{Dominik Kowald}
\author[inst1]{Stefan Thalmann}

\begin{abstract}
aTrain is an open-source and offline tool with an easy-to-use graphical interface for transcribing audio data in multiple languages with CPU and NVIDIA GPUs. It is designed for researchers using qualitative data generated from various forms of speech interactions. aTrain requires no programming skills, runs on most computers, does not require an internet connection, and was verified not to upload data to any server. aTrain combines OpenAI's Whisper model with speaker recognition to provide output that integrates with MAXQDA and ATLAS.ti. It is provided through the Microsoft Store for simple installation. The source code is freely available on GitHub. Having developed aTrain with a focus on speed on local computers, we show that the transcription time on current mobile CPUs is around 2 to 3 times the duration of the audio file using highest-accuracy transcription models. With an entry-level graphics card, transcription speed increases to 20\% of the audio duration.

\end{abstract}




\end{frontmatter}


\newpage
\section{Introduction}
\epigraph{"Transcribing interviews is time-consuming and potentially costly work. It can be facilitated by using a transcribing machine that has a foot pedal and earphones".}{\citep[][p. 118]{seidman2006interviewing}}

In the ten years after the foot pedal was hailed as the state of the art for effectively transcribing interviews, the tools available have developed rapidly with the rise of artificial intelligence (AI). While the transcription of an interview of one hour requires up to six hours of manual work \citep{Bell_Bryman_Harley_2018}, advances in AI-based tools have sped up this process, significantly reducing the necessary transcription work to a fraction compared to manual transcription. An example of such an AI-based transcription model is Whisper, developed by OpenAI in 2023 \citep{whisperpaper}. Whisper uses a transformer model \citep{vaswani2017transformers} to facilitate fast automated transcriptions with accuracy and robustness comparable to human transcribers \citep{whisperpaper}. However, speed and accuracy of the transcription are not the only criteria relevant to qualitative researchers. When comparing the available automated transcription tools, \cite{wollin2023automatic} recently outlined several criteria for the comparison of automatic transcription tools: data protection, accuracy, time spent, and costs. 

Although open-source models such as Whisper perform generally well in all these criteria \citep{wollin2023automatic}, they still lack ease of use and rely on command-line interfaces. This can be a barrier for researchers without programming knowledge, especially when installed locally and operated on their own devices. They also lack output formats that integrate into common QDA software such as MAXQDA or ATLAS.ti. In addition to open-source solutions such as Whisper, there are several paid subscription tools from commercial providers such as Trint, Descript, or Sonix \citep[see e.g.,][]{liyanagunawardena2019automatic, wollin2023automatic} that provide sufficient ease of use for researchers. The problem with some of these existing tools is their cloud-native nature, which requires users to upload their interviews to external servers for transcription. Especially in the context of the EU's data protection legislation (GDPR), this practice of uploading potentially sensitive personal information to cloud services requires explicit consent from interviewees \citep{EuropeanParliament2016}, which they may not be willing to give. Also from the perspective of ethics committees approving qualitative studies, privacy preserving technologies are demanded. 

To help researchers use high-quality open source tools for local interview transcription, we developed a tool providing accessible transcription of interviews: aTrain. 

\begin{figure}[htb]\label{fig_atrain}
\centering
\includegraphics[width=\textwidth]{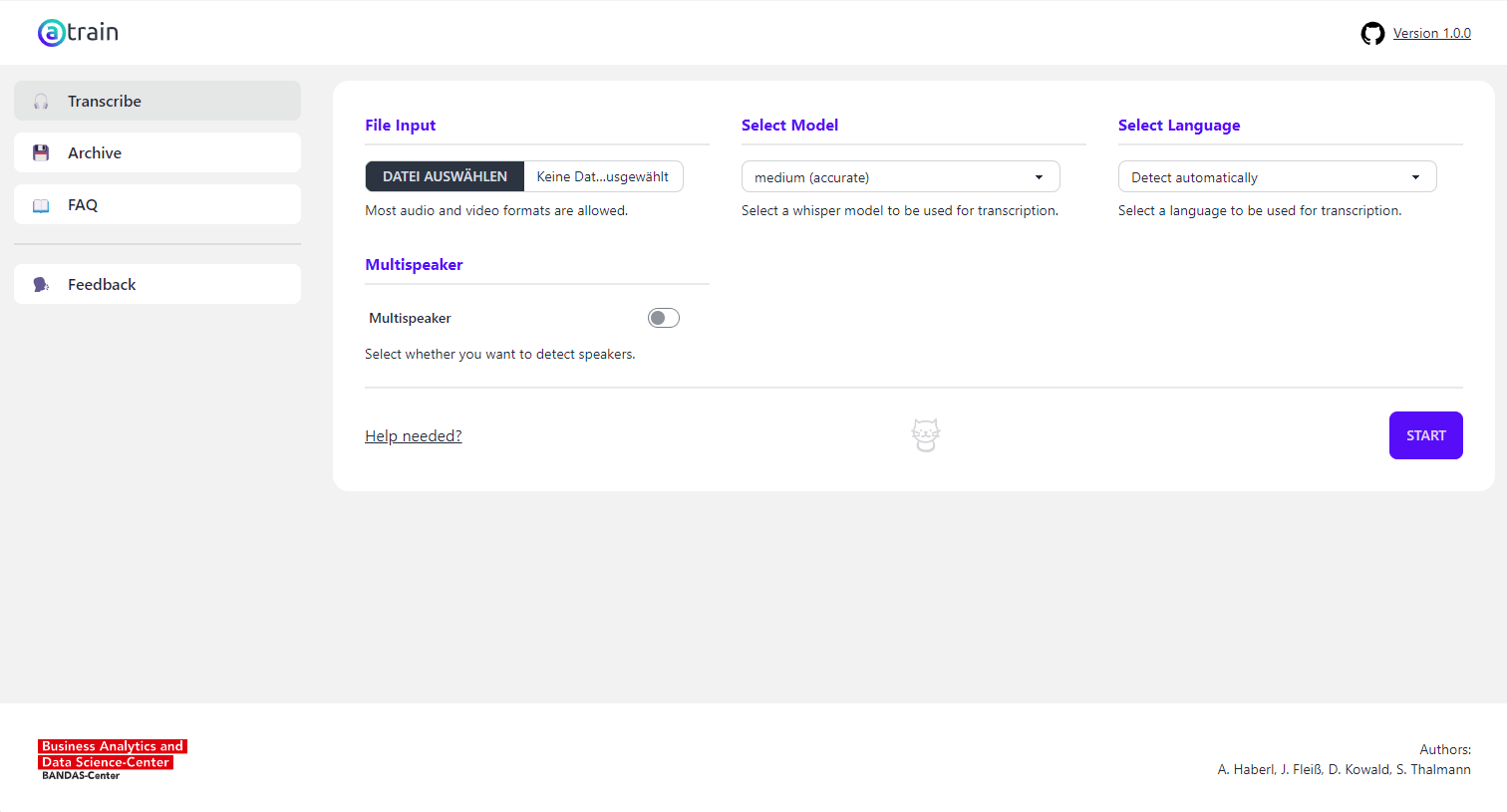}
\caption{aTrain user interface}
\end{figure}

We developed aTrain as a free, open-source, encapsulated, self-installing, and completely offline alternative to existing solutions. The following article presents the main features and details of the technical implementation and benchmarks of the transcription time on different computer configurations. We alleviate three main issues in using free open-source automated transcription on local computers: long runtime, difficulty in setup and use, and integration in the qualitative data analysis software workflow.

To reduce runtime issues and improve on other open source solutions, we make use of the faster-whisper framework, which reduces runtime on CPUs by the factor of 4 to 5 \citep{Klein.2023}, thus making local transcription on typical business notebooks feasible. In our benchmarks, we found that mid-range business notebooks, as are common in university settings, allow one to transcribe one hour of audio with the highest accuracy whisper models in only about 2 and a half to four hours (depending on the hardware). Additionally, aTrain also supports CUDA-enabled NVIDIA graphics cards, drastically reducing the time required to transcribe interviews.

\begin{figure}[tb]\label{fig_maxqda}
\centering
\includegraphics[width=\textwidth]{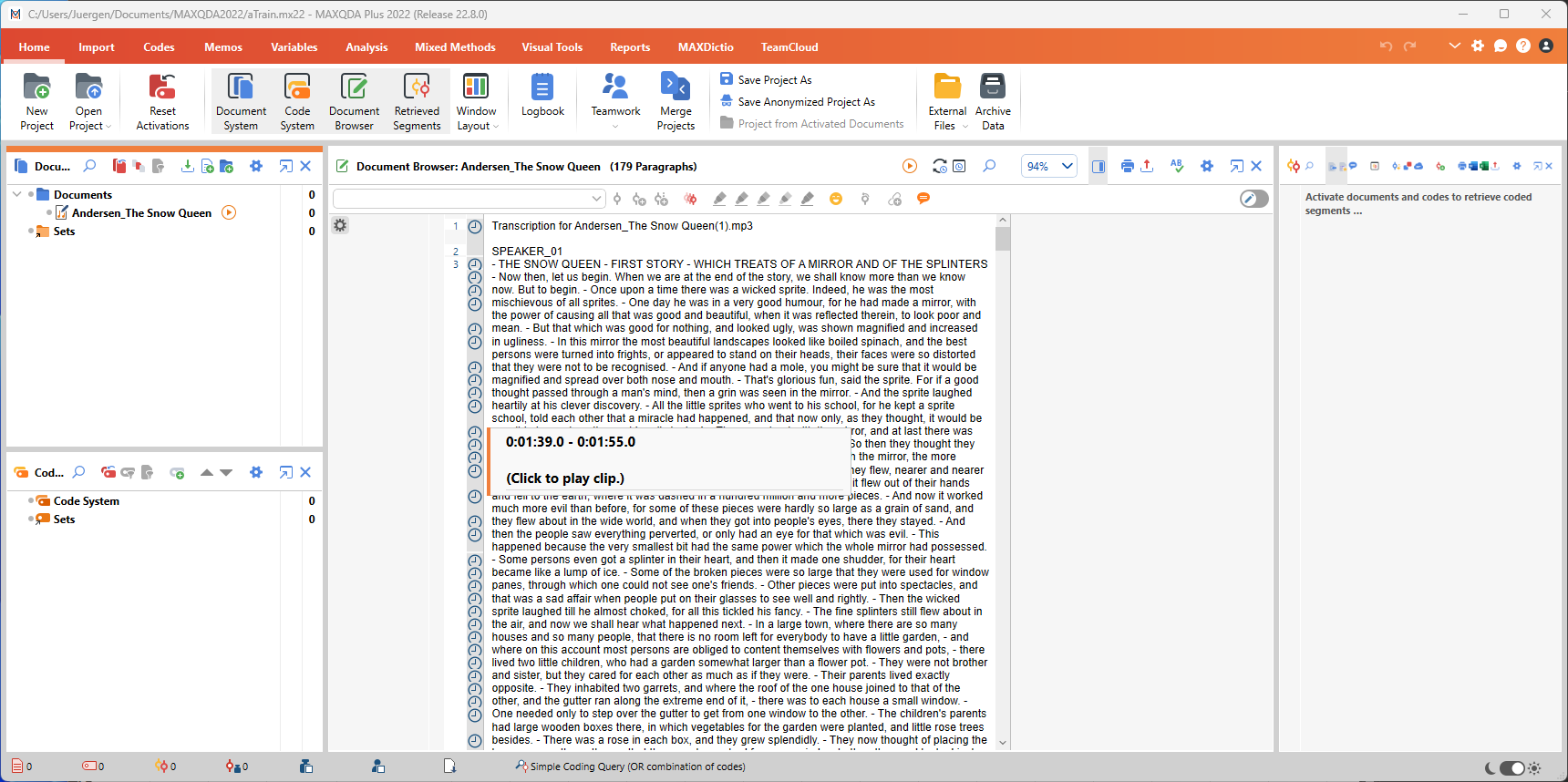}
\caption{Example of a transcript with synced audio in MAXQDA 2022.}
\end{figure}

Regarding ease of use, we offer a Microsoft Store deployment for Windows users and a simple graphical user interface (Figure \ref{fig_atrain}, thus eliminating the barriers of command line setup and use of Whisper. Furthermore, we also offer an export format that allows integration in MAXQDA and ATLAS.ti to allow researchers to continue directly working in established software tools after transcription with aTrain (see Figure \ref{fig_maxqda}).\footnote{To the best of our knowledge, one other open source tool has been developed to alleviate difficulty in setup and use: noScribe. However, the issues of long runtime on CPUs and lack of GPU support have been pointed out as a trade-off that must be made when using this solution \citep{wollin2023automatic}, and that we hope to alleviate.}

Finally, addressing integration in QDA software workflow, we provide an output format for syncing audio with transcript passages in the widely used software tools MAXQDA and ATLAS.ti.

\FloatBarrier

\section{Main Features}
\label{sec:features}

This section provides an overview of the features that the current version of aTrain offers its users. For an in-depth description of the technical implementation of these features, refer to later chapters.

\subsection{Local installation}
To simplify the installation process of aTrain as much as possible, we developed a MSIX package of the application and provide it through the Microsoft Store \citep[see e.g.,][]{microsoft2021MSIX}. The installation does not require administrator rights on the local machine and should even work on systems that are centrally administrated, for example, by university IT departments. aTrain is available through the following link to the Microsoft Store: \newline \url{apps.microsoft.com/store/detail/atrain/9N15Q44SZNS2}.

The source code can be found in the GitHub repository under \newline \url{github.com/BANDAS-Center/aTrain}.

\subsection{Automated transcription}
The central feature of aTrain is its ability to automatically transcribe audio and video files (e.g., mp3, mp4, wav and most others) containing speech.\footnote{For a full list of supported formats, see \url{ffmpeg.org/ffmpeg-formats.html}} After selecting the input file from the local hard drive, aTrain will use its internal ML models to transcribe the contained speech and output the text, including time stamps. There is also an option to specify the ML models which should be used for transcription, which vary in accuracy and speed of transcription. 

It should be noted that in qualitative research, several distinct forms of transcriptions are differentiated, with varying level of detail and the in- and exclusion of phonetic components. A systematic discussion of the different systems and how AI-based translation relates to them is presented by \cite{wollin2023automatic} who conclude that currently all automated systems omit phonetic information. This excludes certain types of qualitative analysis methods that focus on or include phonetic information. Instead, they provide a reproduction of the text approaching a verbatim transcription of what was said, but omitting information like pauses or filler words. However, while the system on which we built our tool, Whisper, was found to be the most accurate and detailed \citep{wollin2023automatic}, it is essential that the researcher reviews the transcript and compares it with the original recording. We highly recommend that researchers using aTrain consult \cite{wollin2023automatic} for a systematic review and discussion of current limitations, capabilities, and advantages of automated transcription tools.

\subsection{Speaker detection}
aTrain additionally offers speaker detection based on Pyannote.Audio \citep[see][]{Bredin.2020}, assigning text passages to the corresponding speakers. Users can either set the number of speakers in a recording, or let aTrain automatically detect the number of speakers. We recommend specifying the speaker count, as we found that it improves the accuracy of the clustering algorithm used for speaker detection.

\subsection{Multi-language Support}
The inputted recordings can contain speech in any of the 57 languages available in the underlying ML models used by aTrain \citep{whisperpaper}. While aTrain can generally transcribe speech in any of these languages, the quality of the transcriptions is generally better for the languages that were predominant in the underlying training data sets\citep{whisperpaper}. Additionally, currently only a single language can be processed for the complete interview.\footnote{One workaround would be to split your audio file into segments based on language manually, using an audio or video editing tool. This would only be feasible, of course, for larger segments with the same language.}

In addition to transcription, translation of audio material into English is also possible by setting the language option to English for non-English source material. Translation to other languages is not supported.

\subsection{Output Formats}
The transcribed text will be exported from aTrain as a text file that contains timestamps for each text segment and optionally information about the corresponding speakers. The following files are provided:
\begin{itemize}
    \item A plain text file containing the transcripts including timestamps and, if selected, speaker information.
    \item A plain text file containing the transcripts without timestamps and, if selected, speaker information.
    \item A version of the plain text file formatted for QDA software import.
    \item A json file containing the complete raw-transcript information, allowing users to construct their own output formats.
\end{itemize}

To ensure easy integration into existing research workflows, we also provide an output format that was designed for seamless import into the commonly used QDA software tools ATLAS.ti and MAXQDA. This then allows to sync the audio or video file to the corresponding points in the transcript, thus playing the corresponding audio or video file by clicking the corresponding text in the transcript.

\subsection{Offline Execution and Data Privacy}

aTrain is designed to run completely offline without the need for an internet connection. Thus, both the transcription model and the data set operate in memory and do not leave the local machine. The same is true for the transcribed text. This means that no data are sent to any server, which ensures data privacy and supports GDPR compliance. 

The version of aTrain referenced in this article and its components have been reviewed with regard to them sharing data online by one of the authors with a programming background. To ensure that updates to the used components do not introduce the transfer of GDPR-related data in updates, we include all components in the provided installer. This results in a large installer size of around 12.8~GB.

\section{Technology and Programming}

\subsection{Hard- and Software Requirements}
\label{sec:hardware}

We tested aTrain on various computer configurations to ensure that it runs on different hardware as well as on more powerful systems with a CUDA-enabled Nvidia GPU.\footnote{For a list of CUDA-enabled GPUs see \url{https://developer.nvidia.com/cuda-gpus}, accessed on 02 October 2023.} While we expect aTrain to also work on similar and less powerful hardware, we do not have data to definitively support this claim.  

aTrain can run on a CPU or on a CUDA-enabled NVIDIA GPU, the latter requiring the installation of the CUDA toolkit \citep{NVIDIA.2023}. While a CUDA-enabled NVIDIA GPU is not required to run aTrain, it significantly improves the speed of transcriptions and speaker detection, as reported in Section \ref{sec_benchmark}. Currently, aTrain is only available for computers running Windows 11. There are no additional software dependencies or requirements.


\subsection{Software Packages and Architecture}

The general architecture of the aTrain desktop application is based on web technologies using the Python programming language and the flask web framework \citep{PalletsProjects.2021}. The user interface was built with HTML, CSS and JavaScript utilizing libraries such as tailwindCSS \citep{TailwindLabs.2023}, daisyUI \citep{Saadeghi.2023}, alpine \citep{Porzio.2023} and htmx \citep{BigSkySoftware.2023}.

\begin{figure}[h!]
\centering
\includegraphics[width=\textwidth]{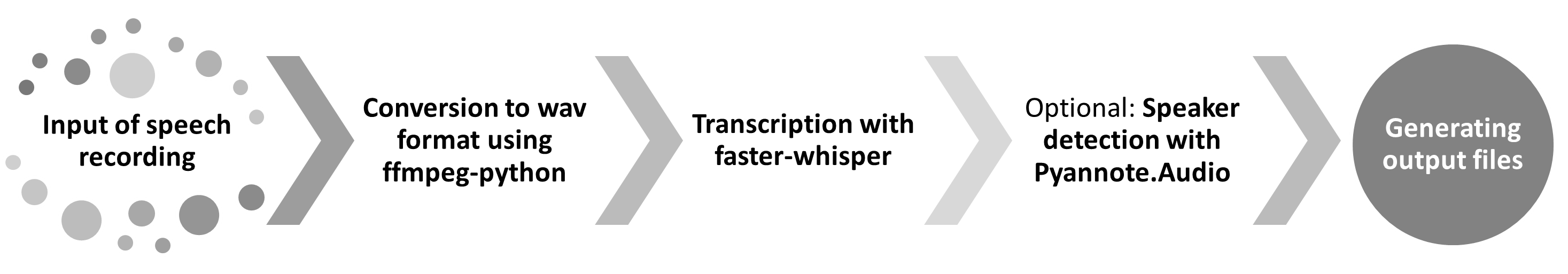}
\caption{Transcription pipeline with corresponding inputs, tools and outputs}
\label{fig_pipeline}
\end{figure}

The main pipeline used for transcription and speaker detection includes several ML models and data processing steps. An important goal in composing this pipeline was to reduce the time needed for transcriptions and speaker detection and to limit the number of ML models included in the final MSIX installer. The final pipeline is depicted in Figure \ref{fig_pipeline}. 

The pipeline is initiated once the user provides a speech recording to the system, which is first converted to the WAV format using a Python binding for the ffmpeg library \citep{Kroenig.2019}. Subsequently, the resulting WAV file is used as input for the Whisper transcription model \citep{whisperpaper} specified by the user. aTrain uses the faster-whisper implementation of Whisper, which is optimized for speed and memory efficiency, which reduces the transcription time on CPUs by a factor of 4 to 5 \citep{Klein.2023}. The output of the Whisper model contains the transcribed text segments of the speech recording, including time stamps. To enable additional speaker detection, aTrain utilizes a modified version of the Pyannote.Audio speaker detection pipeline as described in \cite{Bredin.2020}. An alignment function developed by \cite{bain2022whisperx} is ultimately used to combine the outputs of the transcription and speaker detection models into the final output. 


\subsection{Source code, collaboration and license}
We developed aTrain using Git as the version control system and share the source code through GitHub under \url{github.com/BANDAS-Center/aTrain}. With this, we also want to encourage collaboration with other developers. For future versions of aTrain, we would greatly appreciate the input of other developers and qualitative researchers to expand its features.

aTrain is published under an adaptation of the MIT license, where we ask users to cite this paper when using aTrain for academic or other publications.






\section{Transcription Duration on Different Whisper Models}\label{sec_benchmark}

To enable a comparison of aTrain against existing implementations, we want to provide crucial performance metrics. While we did not develop any ML models ourselves in this project, we did use and combine them in a novel way. Important performance indicators for automatic transcription tools include factors such as transcription accuracy and processing time \citep[see e.g.,][]{wollin2023automatic}. The accuracy of aTrain's transcriptions and speaker detection depends entirely on the underlying models developed by \cite{whisperpaper} and \cite{Bredin.2020}. Accuracy benchmarks and error rates for those models were already documented in the respective papers and are therefore not duplicated in this publication. In general, however, a recent comparison between various commercial and open source solutions found Whisper to be the most accurate tool available, for both a German and an English sample interview \citep{wollin2023automatic}.

A main performance indicator is the total processing time that results from aTrain's transcription and speaker detection pipeline. To test this processing time, we used aTrain to transcribe an audio book which was chosen from the freely available Librispeech corpus \citep{Panayotov.2015}. To emulate the typical length of interviews in qualitative research, we opted to use a recording of Hans Christian Andersen's fairy tail 'The Snow Queen', which was narrated as an audiobook in 1 hour, 13 minutes, and 38 seconds \citep[see][]{LibriVox.2006,Panayotov.2015}.

This audiobook was transcribed with all available Whisper models from the faster-whisper implementation \citep{Klein.2023} and with speaker detection activated \citep{Bredin.2020}. Additionally, we ran the transcriptions on three different computers to test the processing time on different CPUs and also on a CUDA-enabled NVIDIA GPU. The exact specifications of the hardware used during testing and benchmarking are listed in Table~\ref{tab:hardware_spec}. 

\begin{table}[!ht]
\renewcommand{\arraystretch}{3}
\begin{tabular}{ccccc}\label{tab_hardware}
\textbf{Machine} &  \textbf{Year} & \textbf{CPU} & \textbf{RAM} & \textbf{GPU} \\ \hline
\makecell{Dell Latitude \\ 5530} & 2023 & \makecell{\textbf{Intel} \\ \textbf{i5-1245U}} &16GB & \makecell{CPU\\integrated}\\ 
 \makecell{Lenovo Thinkpad \\P14s} &  2022 & 
 \makecell{\textbf{AMD Ryzen} \\ \textbf{7  PRO 6850U}} &  32GB & \makecell{CPU \\ Integrated} \\  
 \makecell{Lenovo Legion \\  Y740} & 2019 & \makecell{Intel \\ i7-8750H} & 16GB & \makecell{\textbf{Nvidia RTX} \\ \textbf{2070 MaxQ 8GB}} \\  \hline
\end{tabular}\\ \\
\caption{Hardware specifications of our experiments. Please note that the computing device used in the benchmarks is highlighted in bold.}
\label{tab:hardware_spec}
\end{table}

The processing time is recorded by aTrain using timestamps generated at the beginning and end of every transcription and displayed as the total duration after the transcription is complete.

\begin{figure}[h!]\label{fig_data}
\centering
\includegraphics[width=\textwidth]{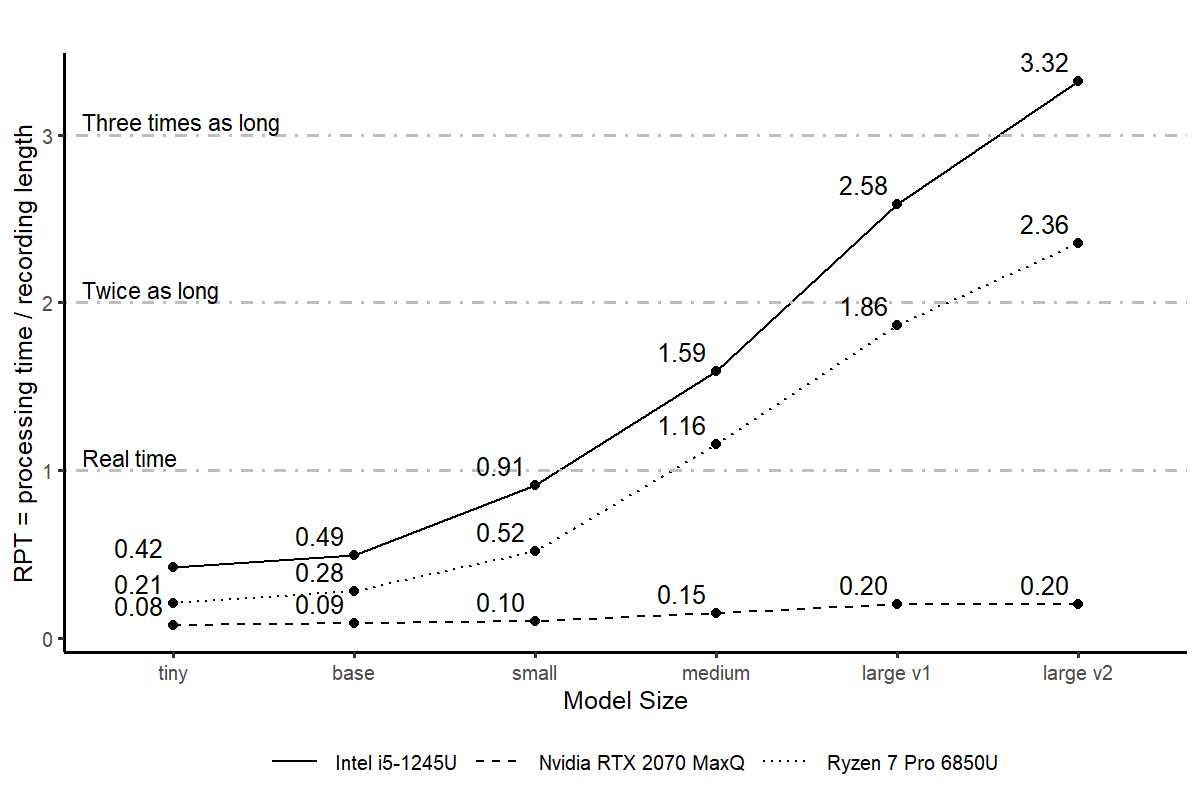}
\caption{Duration of transcription on different hardware}
\end{figure}

Figure \ref{fig_data} shows the processing time of each transcription relative to the length of the speech recording. In this relative processing time (RPT), a transcription is considered 'real time' when the recording length and the processing time are equal. Subsequently, faster transcriptions lead to an RPT below 1 and slower transcriptions to an RPT time above 1.

In alignment with previous expectations, the RPT increases significantly with the size of the Whisper model used. Up to the small Whisper model, the RPT was consistently below 1 for every computer tested. Larger models resulted in RPTs ranging between 1 and 3.5 when running on CPU, while the test machine using a CUDA-enabled NVIDIA GPU (comparable to a 2023 entry-level GPU) could run transcriptions with an RPT around 0.2 even on the largest models. Using standard hardware without GPU (12th gen Intel, Ryzen 7) is nevertheless feasible when running on the medium Whisper model, in which we found a good balance between processing time and transcription quality. However, also the use of large models without a GPU is possible; especially on potent multicore CPUs, where we observed an RPT of around 2 for the large Whisper models. Note that buying an entry-level gaming notebook with a dedicated GPU to transcribe interviews with five times and more real-time speed would cost around the same as a paid automated transcription single-user license starting at around {600\euro} 
 (e.g., \url{https://app.trint.com/plans}, accessed on 3 October 2023). 




\section{Conclusion}
In the paper on hand, we introduce the open-source tool aTrain, which enables the transcription (including speaker recognition) of audio recordings. The software components in aTrain were reviewed by one of the authors with a Computer Science background to ensure that it runs completely on the local computer on which it is installed and does not send any data to online servers, thus helping researchers maintain data privacy requirements arising from ethical guidelines or to comply with legal requirements such as the GDPR.

It comes with an easy-to-use graphical interface, runs on both CPUs and CUDA-enabled NVIDIA GPUs, and is available through the Microsoft Store on Windows computers. aTrain accepts most common audio and video formats as input and produces a transcript for analysis using the Whisper model introduced by OpenAI, providing best-in-class accuracy among current commercial and open-source automated audio transcription solutions. It also provides an output format for import into the QDA software solutions MAXQDA and ATLAS.ti, allowing researchers to play the audio recording corresponding to text passages of the transcript with a single click.

We hope that this tool enables researchers who use audio recordings of various forms of interviews, either as their main data source or as a supplement to more quantitative oriented studies like laboratory experiments, to save both time and monetary resources in transcribing their audio recordings.

\bibliographystyle{elsarticle-harv.bst} 
\bibliography{whisper_paper.bbl}


\end{document}